# How to be correct, lazy and efficient ?

C. Recanati

*Université Paris 13,
av. JB. Clément,
93430, Villetaneuse,
France*

This paper is an introduction to Lambdix, a lazy Lisp interpreter implemented at the Research Laboratory of the University of Paris XI (Laboratoire de Recherche en Informatique, Orsay). Lambdix was devised in the course of an investigation into the relationship between the semantics of programming languages and their implementation; it was used to demonstrate that in the Lisp domain, semantic correctness is consistent with efficiency, contrary to what has often been claimed.

The first part of the paper is an overview of well-known semantic difficulties encountered by Lisp as well as an informal presentation of Lambdix; it is shown that the difficulties which Lisp encouters do not arise in Lambdix. The second part is about efficiency in implementation models. It explains why Lambdix is better suited for lazy evaluation than previous models. The section ends by giving comparative execution time tables.

## 1. Lambdix and the semantic problems of Lisp

In this section the semantical defects of lisp[1] are reviewed and shown not to exist in Lambdix. These defects are characteristic of early versions of lisp, but they are still present in current versions (although, of course, not all defects are present in all versions); this is why we think this little overview is not out of date even if some of points we make are now well-known.

### 1.1. The functional argument problem

Lisp was first thought of as being an implementation of the lambda calculus. Its syntax allows the definition of functions by means of lambda expressions, as for instance

    (lambda (x y) (+ x y))

---

[1] By 'lisp' here we mean a family of languages rather than a particular language belonging to this family.

for addition. This is a fundamental feature of Lisp. Now serious problems arise when these lambda expressions are used as functional arguments - when a function takes another function as argument, and also when a function returns such a function as value.

Example 1
$ ( define  BuildConstFunc (x)
        ( lambda(y)  x ))

The function *BuildConstFunc* is supposed to return a function lambda of y, which returns x. This lambda function ought to be a constant function since its argument y is not used. Then a call to ( (BuildConstFunc 0)  1) ought to return 0, as well as ( (BuildConsFunc 0)  2). This is of course the case in Lambdix, but not in Lisp where:
    $  ((BuildConsFunc  0)   1)
=>  ** error -  x not defined **
or  $  (setq  x  456)
    $  ((BuildConsFunc  0)   1)
    =>  456

The reason for this surprising answer is the following. In most lisps environments (bindings between names and values) are represented in a stack which is popped when the execution of the function terminates. The binding of x to 0 in the application of BuildConsFunc is lost when the function returns the lambda expression; consequently, when the anonymous lambda function is applied, x recovers its previous value - which is 456 or not defined.

Example 2
$  (define apply (f x) (f  x) )
$  ( define  Identity  ( x )
        ( apply (lambda (y)  x) 2))

The *Identity* function defined here is constructed by applying a function returning x. In Lambdix, a call to ( Identity  45)  returns 45 but in lisp:

    $  ( Identity  45)
    => 2

This is even more surprising since only y seems bound to 2. The source of the trouble is the use of the name x in the definition of *apply*. Had *apply* been defined as

    $  ( define apply (f z)
                (f  z) )

the problem would have disappeared. Here it is not because the stack has been popped too early that the binding fails, but because an intermediate binding has been inserted: x is first bound to 45; then the evaluation of *apply* inserts the binding of x to 2. The lambda is evaluated with y bound to 2 and it returns x - which is now bound to 2.

Because of these functional argument problems, Lisp is not a true second order functional language[1], and

$$\$ \ (\ closure\ \ '(x)\ \ 'Expr\ )\ when\ x = 2$$
$$\Rightarrow\ ((lambda\ \ (x)\ \ Expr)\ \ \ '2\ )$$

---

[1] Some lisps proposed a function to solve the functional argument problem. This function, sometimes called *closure*, takes two arguments - a list of formal parameters and an expression - and returns the application of a lambda expression. For instance, the value returned by the *closure* function applied to the list '(x) and an expression Expr - in an environment where x is bound to 2 - is given by :

Although this can punctually solve the problem, many critics can be made to this solution. First, it is not very efficient because it adds function calls and requires the evaluation of all the parameters to be saved. Moreover, it is an ad hoc solution. It requires the programmer to know when the *closure* function is necessary since the call to the *closure* function must be explicit. Furthermore, lisp distinguishing between an f-value and a c-value, some particular attention must be given on the way the interpreter pass the arguments. This requires the use of a special function (called *funcall* ) to force the functional part of the benefits of functional programming is lost.

## 1.2. Lexical scope vs dynamic scope

The foregoing examples show that the names of formal parameters are of major importance in Lisp. This feature is due to the type of variable binding used in Lisp, called 'most recent binding' (MRB, for short). MRB was perhaps, as Gowan has wittily said, the 'most recent error' ([GOW72]). Originally motivated by the technical advantages of the implementation, dynamic scoping of the type illustrated by MRB has disastrous consequences for the safety of the language. *How can you trust a language in which the names of formal parameters must be taken into account?*

The use of dynamic scoping in Lisp conflicts with the original model introduced by Church, which was at the source of Lisp. In lambda calculus, a rule known as the ß-rule allows the reduction of terms along the following pattern:

$$(\ (\ \lambda\ x\ .\ expr\ )\ \ val\ )$$
$$\rightarrow\ \ \ \ expr\ [\ x\ \leftarrow\ val\ ]$$

---

value interpretation and this additional call must also be explicit.

The ß-rule takes as input the application to a value val of a lambda expression with formal parameter x and yields as output the same expression in which all tokens of x have been replaced by val. The substitution is supposed to be determined by the inner lambda binding in the (program) text, i.e. *lexically*.

In most cases, dynamic scope and lexical scope yield the same result, but it is very easy to construct cases with diverging answers[1] . This is why - like many functional languages today (Common Lisp, Scheme, ML, etc.) - Lambdix consistently uses lexical scope for formal parameters. Lambdix also allows the use of functional arguments and is therefore a true second order language.

## 1.3. Evaluation order

### 1.3.1. Call by need and call by value

In lambda calculus, a term t defined by

$(\lambda x\ y\ .\ x)\ A\ ((\lambda u\ .\ u\ u)\ (\lambda u\ .\ u\ u))$

reduces to A because the first reduction (the substitution of A to x) yields a term, $\lambda y\ .\ A$, which always reduces to A because y does not occur in the core of this lambda expression. But in Lisp the interpreter computes the values of the arguments before the core of the function. This strategy of parameter evaluation is known as *call by value*. Since in the evaluation of t all arguments are calculated first, the calculation of the second argument - the term $((\lambda u\ .\ u\ u)\ (\lambda u\ .\ u\ u))$ - generates an infinite loop:

$$y \equiv ((\lambda u\ .\ u\ u)\ (\lambda u\ .\ u\ u))$$
$$\rightarrow ((\lambda u\ .\ u\ u)\ (\lambda u\ .\ u\ u))$$
$$\rightarrow ((\lambda u\ .\ u\ u)\ (\lambda u\ .\ u\ u))$$
$$\rightarrow\ ...$$

In the framework of the lambda calculus, *call by value* can be understood as a strategy governing the order in which the ß-reductions are performed. This strategy guarantees that if there is a unique solution, the calculus will converge on it. Unfortunately it also guarantees that the

---

[1] For instance the term t defined by
t ≡ ( λ x . ( λ y . ( λ x . y) B ) x)   A
would reduced to A by ß-reduction
    t    →    ( λ y . ( λ x . y) B ) A)
        →    ( λ x . A) B )
        →    A
while it would reduced to B in the dynamic model:
( λ x . ( λ y . ( λ x . y) B ) x)   A:

|  | Stack |
|---|---|
| ( λ y . ( λ x . y) B ) x) | [ x – A ] |
| ( λ x . y) B ) | [ x – A ][ y – x ] |
| y | [ x – A ][ y – x ][ x – B ] |
| B | |

A formal demonstration of the non equivalence of the two models can be found in A. Eick and E. Fehr [EIC85].

calculus will never return if there is also an infinite derivation. Thus for terms having both a finite and an infinite derivation, this strategy guarantees that the finite derivation will not be found. Hence it is not a winning strategy. The winning strategy of the lambda-calculus requires that the leftmost inner term be reduced first. If there is a finite solution, the calculus will converge on it (the confluence property guarantees the unicity of the finite solution). This strategy corresponds to *call by need*.

Functions in Lisp are strict[1], because of the strategy of the interpreter (*call by value*). But there is a Lisp function which is not strict: the conditional test. By definition, the *if* function does not calculate all its arguments: the computation of the second and the third argument is determined by the result of the computation of the first. Consequently, nonstrict functions could be constructed if the core of the functions was evaluated before the values of the arguments and if the latter were calculated only when necessary. Such a strategy of evaluation is known as *call by need*. For instance, the function f defined by

( de f (x y)
    (if    (< x 0)
        1
        (f (- x 1) (f x y))))

will never terminate from a call to (f 1 2) if the interpreter conforms to the *call by value* strategy, but it returns 1 if the interpreter conforms go the other strategy (*call by need* ).

The strategic choice made by Lisp is not necessarily objectionable on semantic grounds, because strictness can be seen as a positive feature; moreover, it has the advantage of clarity. But it entails a loss of expressive power, because the set of defined functions is smaller than the set of convergent terms of lambda calculus. The reason why Lisp has made this choice is again efficiency: more often than not, implementations of *call by need* are utterly inefficient.

---

[1] A function is strict if, when applied to an undetermined value, the result is undetermined. This property is usually written by the equation
    $F(\bot) = \bot$
where $\bot$ denotes an undetermined value. In case of several arguments, a function is strict when:
    $F(...,\bot,...) = \bot$

## 1.3.2. Lazy evaluation

An interpreter conforming to the *call by need* strategy is called a *lazy* interpreter. There are various degrees of laziness, however. Pure laziness corresponds to the situation in which the only arguments that are evaluated are the arguments of the printing functions. Such a form of laziness is very inefficient and "lazy evaluation" generally denotes the following pattern:

◊ *call by need* for user defined
      functions
◊ delayed evaluation for the function cons
◊ strict primitive functions
      (+, -, etc.) stand strict

It is the second point which makes the real difference between simple *call by need* for user defined functions and lazy evaluation. The introduction of a lazy constructor on lists sometimes greatly improves efficiency. Lists are the most important structures in Lisp and lazy evaluation offers new ways of handling them.

      Though laziness is a richer model, it has always been considered less efficient than the other evaluation patterns. But is it really less efficient? This question cannot be answered in a straighforward manner, for it all depends on which functions we are talking about. Laziness certainly decreases efficiency in connection with some functions, but it yields fascinating results in connection with other functions. Well exploited, lazy evaluation becomes very efficient in data oriented algorithms, for instance in expert systems or prolog interpreters.

      A lazy cons allows the manipulation of potentially infinite lists called streams. In Lambdix, we can define[1] an infinite sequence of 1 by:

    $ ( de x (cons 1 x))

and nevertheless have

    $ (cadr x)
    = 1

An infinite list of integers can be also defined as a function from:

$ ( de (from x)
    (cons x (from (+ x 1))))

The Lambdix interpreter will have no problem with

    $ (print (cadr (from 2)))
    = 3

---

[1] This is a recursive definition - which is distinct from the traditional setq of lisp:
    $ (setq x 2)
    $ (setq x (cons 1 x))
    $ (cadr x)
    = 2

while Lisp gets trapped into an infinite loop:

= 1 2 3 4 5 6 7 8 9 10 ....

Infinite structures can be very pleasant and efficient in many programs. For instance in numeric application, lazy evaluation is well suited for computing with formal series. The use of infinite structures can also simplify algorithms. It can suppress tests and makes algorithms shorter.
For instance, it suppresses the need for iterators in unification programs.

Stream processing is essential in some simulation programs. It provides an alternative to programming with assignments. The FlipFlop RS gives an interesting example of this feature:

```
$ (de (FlipFlop R S)
      (let ( (de Q (NAnd S QBAR))
             (de QBAR (NAnd R Q)))
        (cons Q QBAR)))
```

Note that the definitions of Q and QBAR are mutually recursive and that the NAnd introduced here will be an operator on streams.

**1.4. Reflexivity in Lisp**

Another difference between Lisp and Lambdix is that Lambdix distinguishes between program and text. Nevertheless, Lambdix provides two operators for an explicit conversion between text and program. This makes it possible for programs to 'modify their own text' during execution.

Primitive objects of Lambdix are typed (numbers, strings, characters, lists, booleans, primitive arithmetic operations, primitive boolean operations, etc.). But there are two levels of language: the level of the program text and that of its semantic interpretation (i.e. its value). In Lambdix *quote* does not indicate the absence of evaluation, as it does in Lisp. *Quote* performs an operation which converts any piece of an already computed part of the program into a list. Lists in Lambdix are constant values; they are not interpreted as forms. A list is a piece of text which remains constant until the operation *excla* is applied. *Excla* is the dual of *quote*. When applied to a list, the list is interpreted in the program text as if the corresponding piece of text had been written there. For instance, the function *mapfun* which constructs the list of the applications of a function f to all the elements of a list l can be written as[1]:

---

[1] An obvious advantage of this definition is that it is perfectly general; there is no need for a distinction between f-subr, f-

```
$ (de (mapfun f l)
     (if (nullist l) ()
         (cons ( ! (cons f (car l)))
     (mapfun f (cdr l))))))
$ (mapfun + '((1 2) (2 3) (3 4)))
   = ( 3 5 7 )
```

Here the characters ! and ' stand for the operators *excla* and *quote* respectively.

## 1.5. Naming variables

A fundamental difference between Lisp and the lambda calculus is that in Lambdix function call is not the only way of binding names to values[1]. Most lisps have at least three other binding mechanisms: affectation (*setq* ), function definition (introduced by *de* ) and local definitions (introduced by *let* ).

### 1.5.1. Naming functions

Names introduced by *de* allow a term to refer to itself in its own definition in such a way that self applicative functions can

---

[1] Without this alternative, the choice of dynamic scoping would really be meaningless

be defined. Top level names in Lisp are mutually defined[2].

Nevertheless Lisp provides mutual definitions only at top level. In Lisp the expression

```
(let ( (var1 expr1)
    ... (varN exprN) )
                  body )
```

is equivalent to the application

```
( (lambda (var1 ... varN) body )
    expr1 ... exprN )
```

ExprN are calculated in the calling environment. Thus cross references cannot be made without additional functional arguments. Since Lisp has problems with second order functions, proper recursive definitions cannot be really introduced at this particular level.

In Lambdix a *recursive* let has been introduced in order to make local

---

[2] This is not a special property of lisp. In the lambda calculus, the fixed point operator allows self reference and then, since function can be pass as arguments, is is possible to write cross referenced terms. In fact, it is lisp which offers restricted possibilities, since functions cannot be passed as arguments (then cross references cannot be done outside the top level definitions).

expr, or macro functions as possible argument values.

function definitions mutual like top-level definitions. The distinction between function value and cell value has also been removed from Lambdix. In Lambdix a term has a unique value. Thus the *de* function can be used for setting any value:

```
$ (de x 3)
= 3
$ (de (f x) (+ x 1))
= f
```

Terms introduced by *de* have mutual references within their own lexical level of definition. The top-level is a special case because top level definitions can be added at any time whereas at other levels definitions come first. Levels can be imbricated. A reference to a name is first looked up at the same level. Parameter names have priority over local definitions when they conflict at the same lexical level[1].

      A recursive let is very useful in a functional language because even if it is possible to handle mutual recursion by means of second order functions, this is quite inefficient and not very clear. Some problems cannot be dealt with without mutual recursion. For instance in the preceding definition of the FlipFlop RS, the terms Q and QBAR must be mutually defined.

### 1.5.2. Naming stores: the assignment problem

Names in Lisp are not used merely as tools for referencing argument values or functions. The basic concept in Lisp is not that of function but that of location. A variable is viewed as a name for a location at which a value can be stored. The set of all bindings at some point in a program is known as the environment at this point. Definitions, lambda expressions and let expressions are viewed as mechanisms which create new locations and bind variables to those locations. Thus formal parameters are not only viewed as potential values but also as stores. In this section we shall consider the problems related to the *setq* instruction.

      First note that *setq* is a *sequential instruction* which must be put inside an enclosing control instruction as *progn*, *while*, etc. This means that function call is not the only way of controlling data flow. This new control level has nothing to do with the lambda calculus - which was our guide up to now.

      The benefit of assignment is efficiency. What is lost is, once again, semantic clarity. The introduction of stores in the semantic model decreases the level of abstraction. For instance, if you

---

[1] If not found as parameter or local definition, the reference is searched in the ancestors.

compare the recursive Fibonacci function definition and the iterative one - based on assignments - it is obvious that the functional definition is very close to the mathematical specification while the other requires a proof.

In Lisp, the top level is an implicit progn in which definitions occur. Definitions are very similar to affectations since they change the values of variables.

Is it possible to introduce a *setq* instruction without radically changing the language? One way of doing so would be to accept the use of *setq* in the body of a let while adding an implicit progn in the core of the let. We could then perform setq instructions on local variables introduced by *let* . This move would not significally change the language since the *let* in question would only mimic the top-level.

To perform assignments on formal parameters, we might add sequentiality through an explicit (or implicit) control instruction in the body of functions. This would obscure the formal specification in embedding a new control level. Nevertheless, if parameters are passed by value, the meaning of the function being defined could not be too much altered. This is not so, however, if assignments on free variables are concerned.

If we accept assignment on global variables, we will loose an interesting property: the fact that the order in which arguments of functions are evaluated does not count. For instance, in the interpretation of:

$ (f (setq x 5) (setq x 6))

the order of evaluation must be taken into account. Note that if a setq instruction is performed within the body of g, we would have the same problem with

$ (f (g 5) (g 7))

In traditional lisps, the problem is solved because the order of evaluation is perfectly determined. In parallel lisps the evaluation order is arbitrary (depending on processors allocation). Thus we must ascertain that in such languages assignments on globals are used only to guide the calculation (for instance to synchronize processes) rather than to contribute to it[1].

In a lazy language, the situation is intermediate. Shared variables are used by only one function at a time and the order of evaluation is determined at run time. Then no special problem of concurrent access arises as with parallelism but the used of global stores must also be carefully specified since the order of evaluation may vary from one execution to another.

---

[1] Global variables are effectively shared and their content can be override by any processes at any time. Hence some more global convention on the way processes may affect these shared variables must be specified.

Another annoying effect of global assignments is that we could obscure the meaning of the function being defined by setting a new value to the name introduced within the body of its own definition[1]. For all these reasons, setq on global variables has not been allowed in standard Lambdix.

## 2. Implementation

The main originality of Lambdix is its implementation model. This section shows that this model can bear comparison with other models on *call by value*, and that it is far better suited for lazy evaluation.

### 2.1. Previous implementation models

---

[1] For instance
(de (f x)
  ....
  (f  (setq f ...))  ... )
To forbid setq on the term being defined would be illusory since the meaning of term relies anyway on the meaning of all other variables defined at the same level.

The original model of Lisp 1.5 implementation used an access variable mechanism called *deep binding*. This model was very efficient with respect to the switching of environments created by function calls, but also fundamentally inadequate to solve the *funarg* problem and totally inappropriate for lazy evaluation. In this model, the cost of a variable access is proportional to the distance (in the tree of dynamic environments) between the current environment and the one where the variable has been bound. Then the cost of a variable access is not bound: the access to a global variable from a recursive call requires a time proportional to the recursion depth.

The MacLisp interpreter has solved this problem by means of a new model called *shallow binding*. This new model was used in most Lisp implementations thereafter. In this model, each variable has a corresponding cell containing its value in any environment. Therefore in any environment the access to a value is direct and constant. Thus variable accesses are very efficient.

Nevertheless on function call, the old values of the argument list must be saved into a stack and the new values stored in the cells so as to reflect the bindings introduced by the function call. Similarly when coming back to the previous environment, the content of the cells corresponding to the argument list must be exchanged with the corresponding stacked values. This makes the binding/unbinding process much more expensive than in the *deep binding* model. So an unexpected consequence of this binding model is that context switching is very expensive.

The original *shallow binding* model was not designed to solve the *funarg* problem. However Interlisp-10 has implemented a solution maintaining a tree of environments instead of a stack. Context switching is done by repeating the binding/unbinding process along the path from the current environment to the future one. A common ancestor must be first determined on the two paths from these environments to the root and then half of the path must be gone through while unbinding the variables, and the second part while rebinding them up to the target environment. Here the cost of context switching is proportional to the bindings between the source environment and the target one. This means that both the number of function calls and the number of arguments are to be taken into account[1].

---

[1] Baker proposed an alternate solution also maintaining a tree of environment. In its solution, the tree is an inverted tree, where the root is always the current environment. To switch from the current environment to another one, one must follow the unique way from the root to this environment and on each node, proceed to the reversal of the pointer and to the exchange of binding values of the environment with the content of their cells. At the end of this process, all identifiers have the correct values in the cells. This model is known as rerooting and an inductive proof of its algorithm is given in [BAC78]. Although this method does not require the search of a common ancestor in the dynamic tree of environments, context switching is still proportional to the bindings from the current environment to the target one and consequently, still very expensive.

What makes an implementation efficient is the cost of variable access and the cost of environment switching. In traditional lisp, the cost of environment switching was not too important because it only concerned neighbour environments and could be handled by a stack mechanism.

Conversely, in second order languages (allowing functional arguments) as well as in lazy languages the cost of environment switching is very important. In both cases, context switching can frequently arise between two distant environments.

In a lazy language, context switches are numerous. This is so because the request for the calculation of an argument occurs within the body of the function, whose environment is by definition different from that of the function call. Therefore, each time a parameter must be evaluated, a context switching must be performed.

If (revised) *shallow binding* is the most efficient of the classical implementation models, it is not very good with respect to laziness. This is so because the cost of a switch between two environments is not limited by a constant but is proportional to the bindings between the environments in the dynamic tree.

## 2.2. Lambdix implementation model

### 2.2.1. Variable access

As in the *shallow binding* schema, each parameter has a cell. But a cell is not attributed to a name, but rather to a formal parameter. The binding of lexical parameter is known in advance by lexical analysis. A reference to a parameter in the core of a function is a direct reference to the corresponding parameter cell. The implementation uses internal structures containing formal parameter cells and the body of the lambda expression. A textual substitution is performed by the function *read* of the interpreter on the body of the function. All names are replaced by the corresponding parameter cells. This substitution[1] corresponds to a partial compilation of lexical references as illustrated by figure 1.

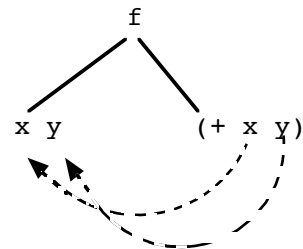

fig. 1

Furthermore, any internal lambda structure contains a pointer to its lexical parent structure, and this substitution is performed at an arbitrary level of definition. For instance, the function double-incr defined by the following top-level definitions

$ ( de  (double-incr x)
            (twice (incr x))))
$ ( de  (twice f)

---

[1] This implementation of the ß-reduction puts Lambdix near the implementation prompted by the De Bruinj notation (like Automath) - where the occurrences of variables were replaced by the level of their binding. But contrary to the De Bruinj model, Lambdix takes care of the names of the variables. This information stands accessible for meta computation. (this allows a certain degree of dynamicity as shown by use of the QUOTE and EXCLA operators).

```
                (lambda (x) (f (f x)))))
$ ( de  (incr  x)
        (lambda (y) (+ y  x)))
```

will be represented by the internal structures of figure 2:

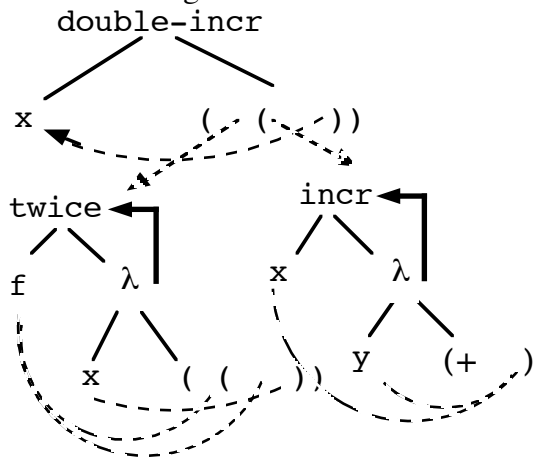

fig. 2

In the same way, local definitions are parsed and lexical references to them are replaced by direct pointers to analogous internal structures. We shall not detail these cases here.

In fact, the cell parameters do not directly contain the values. A pointer to a block is attributed to each function call and the called values are kept into it. Then the cells give access to their corresponding values by means of pointers to the block (called the *dynamic pointers* of the internal lambda structures) and offsets (into the block). This scheme is not as fast as pure *shallow binding* but in this frame variable access is also constant and very efficient. This indirection is interesting because it induces environment switching by changing only dynamic pointers corresponding to function calls. In particular this makes the cost of environment switching independent of the number of parameters introduced by function calls.

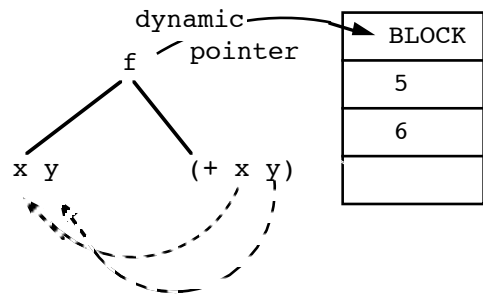

fig. 3

### 2.2.2. Representation of environments

The dynamic pointers to the blocks are used to access values and to represent environments[1]. Environments are

---

[1] The concept of environment generally varies from one implementation to the other. Traditionally environments have been represented as association list - i.e. as functions mapping Identificators to Values:

   sigma: (Id  -> Val)

What we call environment in this paper corresponds to the same notion, while in [REC86], Lambdix environments were introduced by a recursive equation:

   ENV = Id  -> ( ENV  -> Val)

transformed by function calls and function returns. This provides an order on the environments - an order that is usually called 'dynamic'. Athough it is distinct from the lexical ordering of the definitions, they are related. Suppose the functions P, H and G  to be lexically organized with P at the top, H local to P and G local to H. The lambda-structures will reflect this lexical organization by something like fig.4:

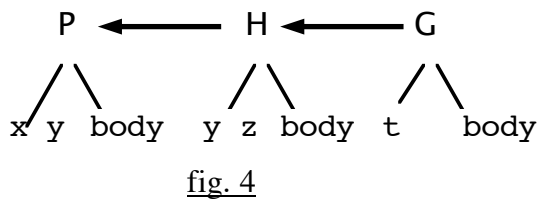

fig. 4

Now a call to H always appears within a call to P. The same thing holds for G and H. The dynamic environment corresponding to a call to G will be defined by the arguments values of this call and those corresponding to its ancestors. In figure 6, the dynamic ordering of the first three calls was: P calling H calling G . Then this first call to G (G1) terminates and a new call to G

These two views are not really opposed; the second one fit better one of the aspects of the implementation model: in a given environment, names give access to function from environments to values and a certain calculation must be performed before accessing a value.

occurs (G2). From this point G calls H (H2) which calls G again (H3).

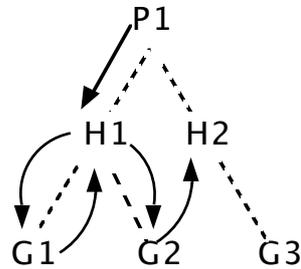

fig. 5

From this last call to G, one must access the parameters of the call (G3) and those of its lexical ancestors (H2 and P1). This block list (G3,H2,P1) combined with the lambda structures G, H, P suffices for representing the variable bindings required during the execution of G3. To install or reinstall this particular environment later, one has only to make the dynamic pointers of the lambda structures G, H and P point to the value blocks G3, H2 and P1 respectively. This is illustrated by figure 6.

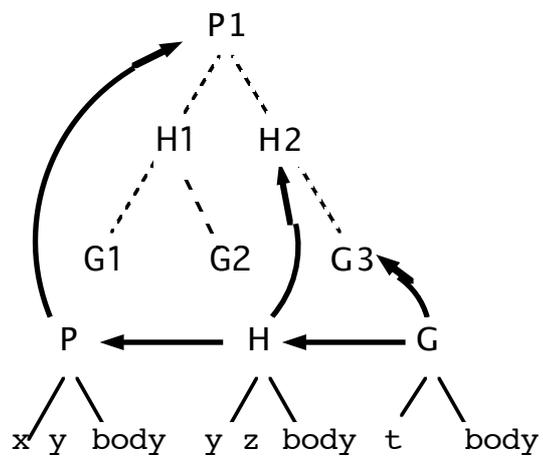

fig. 6

### 2.2.3. Cost of environment switching

In Lambdix implementation, a functional value is represented by a lambda structure and a block corresponding to its environment of definition. This environment corresponds to an environment of call of its direct ancestor. The tree of dynamic environment blocks is maintained by associating each block with its dynamic (lexical) father block. One can then find all dynamic ancestors pointers from one block.

To install an environment of definition on the lambda structures, all ancestor dynamic pointers must be set in the lambda structures to their corresponding blocks in the dynamic tree. To do this, one must simply compare the present values of these dynamic pointers with those given by the tree of calling environments. This calculation can be shortened by the convention that if a pointer is correctly set, all the ancestors will also be. This property can be easily implemented: it only requires that the previous environment be restored when a function returns. In this way the cost of environment switching is limited to one assignment and one test when a function call occurs within the same branch of the lexical tree.

Now the switching from one environment to an arbitrary other (case of closures given as functional argument) will also be limited because it will stop as soon as an ancestor pointer is correctly set - which means that in the worst case, it requires the comparison of all dynamic pointers to the top-level. But here, the number of tests and assignments involved is limited by the lexical level of the definition. It is not a dynamic depth that is involved in this process. It is a lexical depth, which corresponds to the number of levels introduced by let - which usually reduces to 1.

What is important is that the cost of the installation of an environment does not depend on the current environment since it is limited by a value that is independent. It does not matter from which environment the interpreter comes, but how deep this environment is in the lexical sense. Furthermore, if the environment of definition is the same as the previous environment (as for instance in recursive call) environment switching is reduced to a test.

In the general case, the cost of environment switching does not depend on the number of formal parameters, as in the *shallow binding* schema. Thus, functions of multiple arguments are not disadvantaged.

The combination of these two characteristics — constant variable access and environment switching independent of the current environment — makes our model well suited to lazy evaluation because the computation can be postponed without too drastic a supplementary cost.

### 2.3. Execution timetables

In this section we give some execution timetables, although everyone knows that such results cannot not be taken too seriously.

This is the case, first of all, because languages have different semantics. This means that some functions defined in one language may be not defined in another. This situation arises for Lambdix both with second order functions and lazy functions manipulating infinite lists[1]. It is nevertheless possible to compare execution times of those functions that can be defined in both languages.

Secondly, if we want to compare two models for the implementation of environments, we should compare interpreters written in the same implementation language and as far as possible by the same programmer; otherwise, we will not know if the results are due to the efficiency of the implementation language, to the programmer's ability, or to the model itself.

To test our implementation model, we have built two Lambdix interpreters: the first supporting only *call by value* and the second performing only *call by need*. Both interpreters are built on the same implementation schema.

### 2.3.1. Performances on call by value

We compare Lambdix with lisps that are the most used at the L.R.I.: last Lisp (called lal ), Frantz Lisp and Lelisp. These lisps all suffer from the semantical defects listed in the first section[2].

---

[1] We only have compared the very same programs. We did not take advantage of possible simplification of lazy program.

[2] Originally, a team has developed software in Frantz Lisp because it was used in the industry and because it was the only one to have a compiler. Unfortunately, the interpreter was very slow because its implementation was based on the deep binding model. Thus Last lisp has been designed to be an efficient interpreter compatible with Frantz Lisp. Lelisp has been chosen by another team for the efficiency of its interpreter and its good library (which by the way contains a *closure* function). implementation is based on the *shallow binding* schema.

Last Lisp was written in C at the L.R.I. in 1986 by a very good programmer, Patrick Amar, and the interpreter is optimized by assembly code insertions. The LeLisp interpreter was directly written in Assembler. Given that our implementation is presumably less optimized than those of its two competitors, the results[1] shown in the following table look very good for the Lambdix environment model:

| Prog  | lal  | Lambdix | lelisp |
|-------|------|---------|--------|
| Tak   | **38.5** | 42.5 | 40 |
| Fib   | **8.8**  | 13.8 | 13.8 |
| LComp | 15.5 | 11.1 | **9.1** |
| Sieve | 9.1  | 8.1  | **6.5** |

fig. 7

The Fibonacci test (Fib) is executed with a call to (Fib 20). The well known Tak function is tested with a call to (Tak 18 12 6). The last two tests are small programs containing recursive calls to functions which manipulate lists. LComp is a program which compares trees leaf by leaf (same fringe problem). Sieve constructs the list of the first 400 prime numbers with the traditional algorithm of Eratosthenes.

---

[1] The experimentation has been done on a VAX 750, and time is expressed in second. To have the equivalent order on a Sun3.5 you must divided these numbers by a factor 2.

Even if it is never the best (on *call by value*), our interpreter can bear comparison with its two rivals. It gives better results than last Lisp on functions using lists and is close to LeLisp on recursive functions manipulating numbers. What is shown by these results is clearly that the Lambdix implementation model can compete with the *shallow binding* model.

Comparison with the Frantz Lisp interpreter and compiler (fig. 8) illustrates two other points:

◊ *deep binding* is not an efficient model
◊ good interpreters can compete with compilers:

| Prog  | best | Frantz Lisp | |
|-------|------|---------|--------|
|       |      | interp. | comp.  |
| Tak   | 38.5 | overfl. | 37 |
| Fib   | 8.8  | 108 | 6 |
| LComp | 9.1  | overfl. | 10 |
| Sieve | 6.5  | overfl. | 5.5 |

fig. 8

### 2.3.2. Efficiency vs laziness

We have built two Lambdix interpreters: the first one supporting only *call by value* and the second one performing only *call by need*. The comparison of these two interpreters is surely very instructive because the two implementations both share the same implementation language, the same model (adapted for lazy evaluation) and the same programmer. Figure 9 gives a comparison between Lambdix with *call by need* and Lambdix with *call by value*. We have not shown the performances of other lisps because they can be found in the previous tables and are in any case of the same order as Lambdix with *call by value*.

| Prog | Lambdix by val | Lambdix by need | % of diff. |
|---|---|---|---|
| Fib | 13.8 | 15.7 | - 12 |
| Fib2 | 19.5 | 21.7 | - 10 |
| Tak | 42.5 | 57 | - 25 |
| Comp | 11.1 | **0.7** | **+ 94** |
| Sieve | 8.1 | 9.5 | - 15 |
| LSum | 16.1 | **0.03** | **+ 99** |

fig. 9

The programs tested in this table are the same as the previous ones except for Fib2 and LSum. Fib2 is a modified Fibonacci function which takes three arguments[1]. LSum calculates the sum of the terms of two lists generated by the Sieve program.

---

[1] Other lisps were very sensible to these additional arguments and gave all a result around 20. Note that this is not the case with the Lambdix model which is less sensible to parameter addition.

Lazy evaluation gives better results because the sum is computed only on the first elements.

As shown by this table, the bad cases of lazy evaluation are between 10% and 25% slower than traditional *call by need* while the good cases are 95% better. The worst case is that of the Tak function which has three arguments and calculates them separately - each time from another environment. But such cases are not very frequent and the price of laziness can be evaluated in our model as a 15% loss in efficiency with strict functions. Such a result is very important and could not have been obtained with other models because of the cost of environment switching.

---